\pdfoutput=1

\documentclass[twocolumn,aps,prl,showpacs,superscriptaddress]{revtex4-1}
\usepackage{graphicx}
\usepackage{amsfonts}
\usepackage{amsmath}
\usepackage{amssymb}
\usepackage[colorlinks=true,citecolor=blue,urlcolor=blue]{hyperref}

\def\endproof{\vrule height6pt width6pt depth0pt}

\def\be{\begin{equation}}
\def\ee{\end{equation}}

\begin{document}


\title{Quantum theory allows for absolute maximal contextuality}


\author{Barbara Amaral}
\email{barbaraamaral@gmail.com}
 \affiliation{Departamento de Matem\'atica, Universidade Federal de Ouro Preto,
 Ouro Preto, Minas Gerais, Brazil}

\author{Marcelo Terra Cunha}
\email{tcunha@ime.unicamp.br}
 \affiliation{Departamento de Matem\'atica, Universidade Federal de Minas Gerais,
 Caixa Postal 702, 30123-970, Belo Horizonte, Minas Gerais, Brazil}
 \affiliation{Departamento de Matem\'atica Aplicada, IMECC-Unicamp, 13084-970, Campinas, S\~{a}o Paulo, Brazil}

\author{Ad\'an Cabello}
\email{adan@us.es}
 \affiliation{Departamento de F\'{\i}sica
 Aplicada II, Universidad de Sevilla, E-41012 Sevilla, Spain}
 \affiliation{Universidade Federal de Minas Gerais,
 Caixa Postal 702, 30123-970, Belo Horizonte, Minas Gerais, Brazil}


\date{\today}



\begin{abstract}
Contextuality is a fundamental feature of quantum theory and a necessary resource for quantum computation and communication. It is therefore important to investigate how large contextuality can be in quantum theory.
Linear contextuality witnesses can be expressed as a sum $S$ of $n$ probabilities, and the independence number $\alpha$ and the Tsirelson-like number $\vartheta$ of the corresponding exclusivity graph are, respectively, the maximum of $S$ for noncontextual theories and for the theory under consideration. A theory allows for absolute maximal contextuality if it has scenarios in which $\vartheta/\alpha$ approaches $n$. Here we show that quantum theory allows for absolute maximal contextuality despite what is suggested by the examination of the quantum violations of Bell and noncontextuality inequalities considered in the past. Our proof is not constructive and does not single out explicit scenarios. Nevertheless, we identify scenarios in which quantum theory allows for almost absolute maximal contextuality.
\end{abstract}



\maketitle


\section{Introduction}


Contextuality (namely, the impossibility of explaining probabilities of measurement outcomes as originated from pre-existent values which are not altered by compatible measurements \cite{Specker60,Bell66,KS67}), has recently been identified as a critical resource for quantum computing \cite{HWVE14, Raussendorf13, DGBR14} and, through nonlocality (a form of contextuality), device-independent secure communication \cite{Ekert91,BHK05}.
Recent progress has allowed us to identify where, how, and why quantum contextuality occurs. In particular, recent results comprise a necessary and sufficient condition for quantum contextuality \cite{CSW10,CSW14}, necessary conditions for quantum state-independent contextuality \cite{RH14,CKB15}, the maximum contextuality possible for any structure of exclusivity \cite{CSW10,CSW14}, and a number of principles that explain the quantum limits of contextuality for certain important scenarios \cite{PPKSWZ09,NW10,Cabello13,FSABCLA12,Yan13,ATC14,Amaral14,Cabello14,Cabello15}.
In addition, several quantifiers of contextuality have been introduced \cite{GHHHJKW14}, the connections between contextuality, entanglement, and nonlocality have been explored \cite{Cabello10,KCK14}, and the relationship between quantum contextuality and maximally epistemic interpretations of quantum theory (QT) has been examined \cite{LM13}.

Still, we know very little about the contextuality that can be produced with quantum systems. For example, is QT the most contextual theory possible? Initially, despite a very appealing candidate as an explanation for QT, the examination of isolated scenarios suggests that the answer should be negative \cite{PR94} and that QT is ``neither the most nonlocal theory one can imagine, nor the most contextual'' \cite{CY14}. However, when one applies some simple principles to copies of the scenario \cite{PPKSWZ09,NW10,Cabello13,FSABCLA12} or to extended scenarios which include extra possible measurements \cite{Yan13,ATC14,Amaral14,Cabello14,Cabello15}, one finds out that, at least in some key cases, the maximal quantum contextuality of the original scenario is restricted by the fact that this scenario can be embedded into a larger one which is as contextual as possible (assuming some of these simple principles).
This suggests that QT is the most contextual theory allowed by some principles. However, this leads to another question of fundamental and potentially practical implications: How much contextuality is that? How large contextuality can be in QT? In which experiments does it occur? How does the ratio between the maximal allowed contextuality and the noncontextual bound behave? How does this maximum quantum contextuality compare with the contextuality allowed by other theories not restricted by the principles that limit quantum contextuality?

In this paper we address these questions. {\em A priori}, they are difficult questions. To deal with them, we first observe that linear contextuality witnesses with all coefficients equal to one can be written as a sum $S$ of $n$ probabilities such that $\alpha$ and $\vartheta > \alpha$ are, respectively, the maximum of $S$ for noncontextual theories and for the theory under consideration. Clearly, $1 \le \alpha < \vartheta \le n$. This motivates the following definition: A theory allows for absolute maximal contextuality (AMC) if it has (at least one) family of experimental scenarios in which $\vartheta/\alpha$ approaches $n$. As defined in the abstract, $\alpha$ is a well known graph theoretical invariant, the independence number.
For QT, the Tsirelson-like bound \cite{Tsirelson93}, $\vartheta$, also corresponds to a well known graph theoretical invariant \cite{CSW10,CSW14}, the so called Lov\'asz number \cite{Lovasz79}. It is interesting to stress that from the complexity theory viewpoint, the quantum bound is an ``easy'' problem (given a graph $G$, $\vartheta$ is the solution of a semidefinite program), while the noncontextual $\alpha$ is a nondeterministic-polynomial-hard (NP-hard) problem \cite{Karp72}. This identification gives a systematic way of exploring the $\vartheta/\alpha$ ratio for QT. We explore its predictions for $n \le 10$ and examine well-known Bell inequalities to get some insight on the behavior of $\vartheta/\alpha$ in QT. This exploration suggests that QT does not allow for AMC. We then present the main result: QT does allow for AMC. Our proof, however, is not constructive and does not identify specific scenarios. To ease this lack, we present quantum scenarios with almost AMC. We end by discussing why AMC is actually possible in QT and suggest that AMC is an emerging phenomenon.


\section{Contextuality witnesses and graphs}


The question of how large contextuality can be in QT is very difficult to answer if one adopts the traditional approach within which most measures of contextuality \cite{GHHHJKW14} are defined. By he traditional approach, we mean that which starts with a previously specified experimental scenario (i.e., a number of observables and their relations of co-measurability) in which we have to obtain the noncontextuality (NC) inequalities and compute their maximum quantum violations (see, e.g., Ref.~\cite{AQBTC13}). The difficulty is that the number of scenarios is infinite, the number of NC inequalities grows enormously with the number of observables, and the computation of the quantum maxima becomes unfeasible even for relatively simple scenarios.

Interestingly, the problem can be addressed by adopting the graph-theoretic approach to quantum contextuality \cite{CSW10,CSW14,AFLS12}. In this approach, any quantum (linear) contextuality witness that can be expressed as a finite sum of probabilities, i.e., any NC inequality involving a linear combination of probabilities, can be ascribed to a graph with certain properties and, reciprocally, from any graph with these properties one can obtain a quantum contextuality witness of this type. More precisely, for any given graph $G$, there is always one quantum experiment such that the noncontextual bound is given by the independence number of $G$, $\alpha(G)$ (i.e., the maximum number of nonadjacent vertices in $G$), while the maximum value in QT is given by the Lov\'asz number of $G$, $\vartheta(G)$ \cite{Lovasz79}. It is valuable to comment on the generality of this approach. Any tight contextuality witness that is linear in terms of probabilities can be written as a sum $S$ of probabilities. The reasoning is two-fold: First, one can eliminate negative coefficients by changing the probability appearing in each corresponding term by one minus the complementary probabilities; second, given that all extremal points in the noncontextual set have integer coordinates, the corresponding facets can be written with rational (and hence integer) coefficients. The only kind of contextuality witness that is disregarded by this approach is the nonlinear one.


\begin{figure}[tb]
 \centering
 \includegraphics[width=0.38 \textwidth]{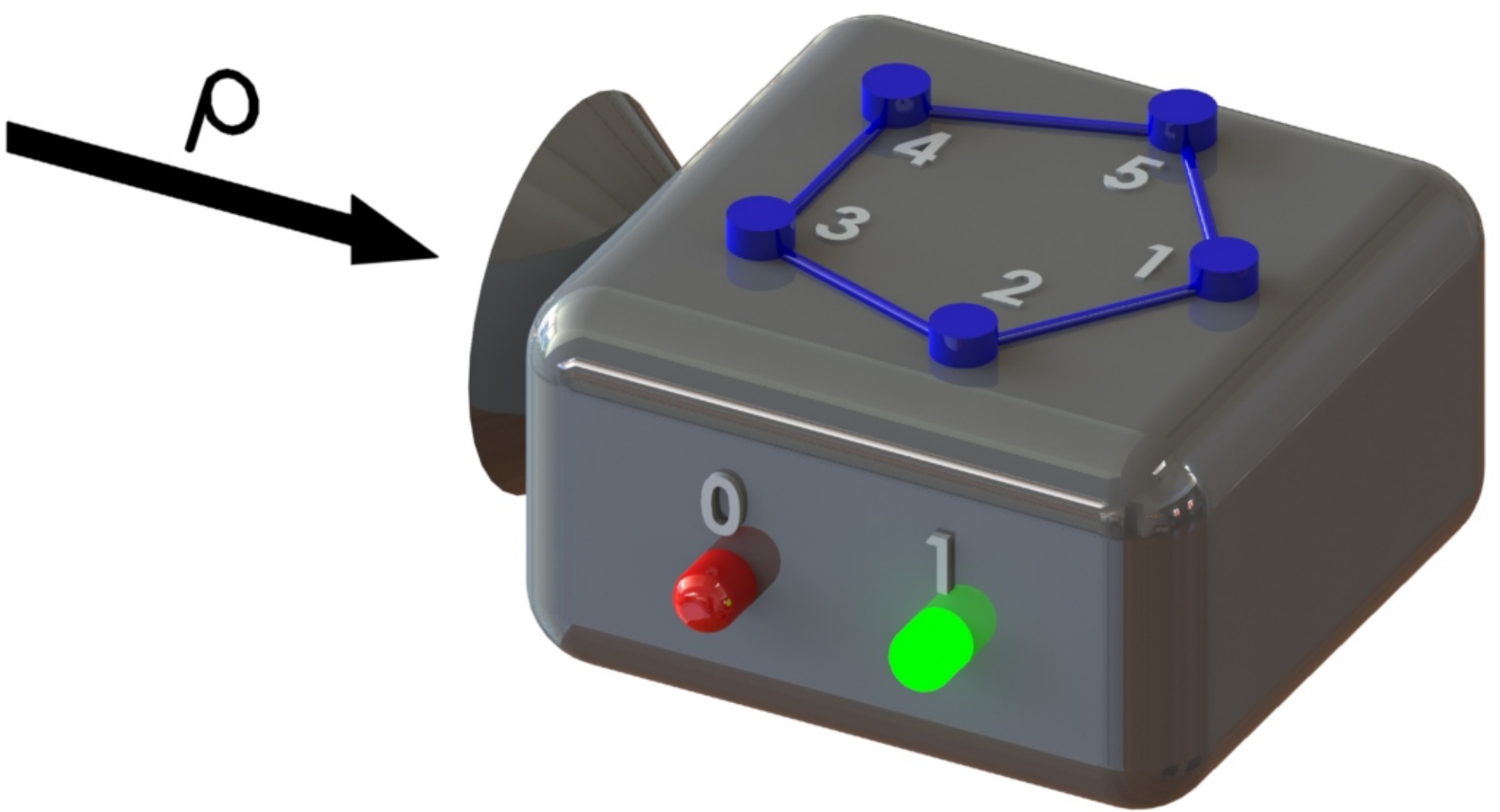}
 \caption{(Color online) A physical system in state $\rho$ enters the box. The player presses a button $i$ (with $i=1,\ldots,5$) and a light ($r=0,1$) flashes. The player can also press a second button $j$ such that $i$ and $j$ are adjacent in the graph drawn at the top of the box. Two important properties can be checked: (i) adjacent buttons correspond to tests without mutual disturbance, i.e., $P(i=r|ij)=P(i=r|ik)$ for any $j,k$ adjacent to $i$; (ii) exclusiveness relation: no adjacent vertices can ``happen'' together, i.e., $P(i=1,j=1|ij)=0$, for all $i.j$ adjacent. \label{Fig1}}
\end{figure}


The ratio $\vartheta(G)/\alpha(G)$ is a natural measure of quantum contextuality within the graph approach. In addition, the fact that $\vartheta(G)/\alpha(G)$ is a good measure of contextuality can also be justified by appealing to the following betting game (see also the game proposed in Ref.~\cite{CY15}). Consider a bookmaker, Bob, that accepts all kinds of bets. A gambler, Alice, brings him a preparation device and a box with a graph $G$ drawn at the top and whose properties are described in Fig.~\ref{Fig1}. The bookmaker can check that both the preparation device and the box work as promised.

At each run, Alice chooses one of the buttons and bets $c$ units of money that pressing this button will give the result $1$. The gain $g_i$ is defined by Bob agreeing to pay her ${c}{(g_i - \epsilon)}$ units of money, where $\epsilon > 0$.
As a bookmaker, Bob establishes $g_i$ in order to guarantee his profit after many rounds of the game and to make the betting attractive to Alice. Denoting her betting probability for button $i$ as $b_i$, the expectation of the pay-off is ${\sum_i b_i P_i (g_i - \epsilon )}$, where $P_i = p(i=1|i)$ is the probability that pressing the button $i$ flashes the green light $1$.
If Bob believes that the system is noncontextual, then he will estimate the prize trusting that ${\sum_i P_i \leq \alpha(G)}$. His simplest choice is to estimate the gain as $g_i = [\alpha(G) b_i]^{-1}$, which gives the expected pay-off as $\displaystyle{\frac{1}{\alpha(G)}\sum_i P_i - \epsilon \sum_i b_i P_i}$, which Bob trusts to be less than $1$.
For a quantum gambler, however, $\sum_i P_i$ can reach $\vartheta(G)$. This means that a quantum gambler playing against a noncontextual bookmaker is expected to make a profit about $\displaystyle{\frac{\vartheta(G)}{\alpha(G)} - 1}$ per unit of money, per round (in the limit $\epsilon \rightarrow 0$).


\section{Is there any indication of absolute maximal contextuality in quantum theory?}


Here we review what is known about the quantum maximum of $\vartheta/\alpha$ for NC and Bell inequalities.

For contextuality witness $S$ defined as a sum of $n$ probabilities, the quantum maximum of $\vartheta/\alpha$ is known for any $n \le 10$ \cite{ADLPBC12}. These quantum contextuality witnesses exist only if $n \ge 5$ \cite{CSW10,CSW14}.
Then the quantum maxima of $\vartheta/\alpha$ are $\sqrt{5}/2 \approx 1.118$ for $n=5,6,7$, $2 (2 - \sqrt{2}) \approx 1.172$ for $n=8$, $11/9 \approx 1.222$ for $n=9$, and $5/4=1.25$ for $n=10$ \cite{ADLPBC12}. Still, any of these maxima is very far from the values required for the AMC. For explicit families of NC inequalities with increasing number of settings \cite{AQBTC13}, the distance to the AMC is indeed growing with the number of settings.

Bell inequalities violated by QT are also quantum contextuality witnesses. The advantage with respect to NC inequalities is that, unlike NC inequalities, Bell inequalities have been extensively studied for years and many results and examples for the growth of the ratio $R$ between the quantum and noncontextual (i.e., local) bounds are known, because this ratio usually measures the quantum vs classical advantage for certain tasks involving separated parties. Here we present a brief overview of such results (see Ref.~\cite{Palazuelos15} for more details) and explain how these results are connected to our problem. For our purposes, $R$ will play the role of $\vartheta/\alpha$ and, hereafter, $m$ will be the number of parties, $d$ the dimension (of at least one) of the local subsystems, $N$ the number of settings per party, and $K$ the number of outputs, while $n$ is the number of probabilities in the contextuality witness $S$.

In a pioneering work, Tsirelson \cite{Tsirelson93} showed that, for bipartite Bell inequalities, $R$ is upper bounded by the real Grothendieck constant $K_G^\mathbb{R}$ \cite{Grothendieck53}, whose exact value is unknown (but bounds are known). In the multipartite case, the first investigations focused on $R$ for particularly promising quantum states. For example, for tripartite Greenberger-Horne-Zeilinger states, $R \le 4\sqrt{2}K_G^\mathbb{C}$, where $K_G^\mathbb{C}$ is the complex Grothendiek constant (whose exact value is also unknown). For Schmidt states, $R \le 2^{\frac{3m-5}{2}}K_G^\mathbb{C}$ and $n \sim N^m$. Similar results hold for clique-wise entangled states and, in particular, for stabilizer states \cite{Palazuelos15,JPPVW10,JP11}.

Better results where found with the help of the probabilistic method. In the tripartite scenario it is possible to prove, for every $d \in \mathbb{N}$, the existence of quantum violations with $R \sim O(\sqrt{d})$, with $n \sim O(2^{d^2}\times 2^{D^2} \times 2^{D^2})$ \cite{PWPVJ08}. However, this result is highly nonconstructive and there is no estimate for $D$ (the dimension of the other two subsystems). In Ref.~\cite{BV13} the authors show that there are tripartite Bell inequalities for which $R \ge c\sqrt{N}\log^{-\frac{5}{2}}N$, for some constant $c>0$, while $n \sim O(N^6)$, improving the previous result.

There is a general result which provides an upper bound for $R$ for tripartite correlation Bell inequalities. In Refs.~\cite{PWPVJ08, BV13} it is proven that, for this case, $R \le O(\sqrt{k})$, where $k=\min\{N,d\}$. In this case, $n \sim O(N^3)$. A generalization of this result to $m$ parties shows that $R \le O(N^{\frac{m-2}{2}})$, with $N$ settings for at least $m-2$ parties \cite{BV13}. In this case, $n \sim O(N^{m-2})$. Both results are also nonconstructive and there are no explicit examples approaching these bounds. For these nonconstructive results, $n$ is a loose estimative, since many Bell inequalities involve only a fraction of the $K^{m N}$ probabilities.

For a large number $m$ of parties, large violations can be obtained already in the simpler scenario with two settings per party. For the inequalities connected with XOR games, which include the Werner-Wolf-{\.Z}ukowski-Brukner inequalities \cite{WW01, ZB02}, $R=2^{\frac{m}{2}}$ can be found for some particular inequalities, including the Mermin inequalities \cite{Mermin90, Ardehali92}. This value is optimal for this scenario. In this case, $n \sim O(2^m)$.

For general bipartite inequalities, $R \le O(h)$, with $h=\min\{N,K,d\}$ \cite{JP11}. In this case, $n \sim O(N^2 \times K^2)$. If $N=K=d$, there are Bell inequalities with $R \ge \Omega\left(\frac{\sqrt{N}}{\sqrt{\log N}}\right)$. In this case, $n \sim O(N^{4})$ \cite{JP11}. In Ref.~\cite{JPPVW10} the authors prove the existence of Bell inequalities with
$R \ge \frac{\sqrt{k}}{\log^2 k}$, where $N=\left(2^{\frac{\log^2 k}{2}}\right)^k$. There are no explicit examples achieving these bounds. Also in these cases large violations can be obtained but at the expense of increasing exponentially $n$.

In the case of general bipartite Bell inequalities, two important explicit examples are shown in Ref.~\cite{BRSW11}. The first one is the family of inequalities associated with the hidden matching game, which has $R \sim O(\frac{\sqrt{K}}{\log K})$ for $N=2^K$, which gives a superexponential $n$. The second example is the family of inequalities associated with the Khot-Vishnoi game, with $R \ge \Omega\left(\frac{\sqrt{K}}{\log^2 K}\right)$, with $N=\frac{2^K}{K}$, which also gives a superexponential $n$. To our knowledge, the growth of $R$ for Bell inequalities with three or more parties remains unexplored.

As we see, none of these quantum violations of NC or Bell inequalities even remotely suggest that QT might allow for AMC.


\section{Absolute maximal contextuality in quantum theory}


Despite what is suggested by all previous evidence, the following theorem holds:

\emph{Theorem 1.} Quantum theory allows for absolute maximal contextuality.

\emph{Proof.} The proof is based on two previous results. In Ref.~\cite{CSW14} it is shown that, for any $n$-vertex graph $G$ such that $\alpha(G)<\vartheta(G)$, there is a quantum contextuality witness $S$ such that $\alpha(G)$ and $\vartheta(G)$ are, respectively, the noncontextual and quantum tight maximum of $S$. Its physical implementation requires us to prepare a quantum state in the handle of a Lov\'asz-optimum orthogonal representation of the complement of the graph and to measure the rank-one projectors onto the unit vectors of that representation (see Ref.~\cite{CSW14} for details). Therefore, proving the existence of scenarios of AMC requires proving the existence of graphs such that $\vartheta(G)/\alpha(G)$ approaches $n$. Reference \cite{Feige95} proves that, for every $\epsilon > 0$, an $n$-vertex graph $G$ exists such that $\vartheta(G)/\alpha(G) > n^{1-\epsilon}$. \hfill \endproof

The proof in Ref.~\cite{Feige95} uses the probabilistic method and no explicit construction approaching these values is known. Therefore, although the existence of scenarios allowing for AMC in QT is guaranteed by the result in Ref.~\cite{CSW14}, we cannot present any explicit example. To ease this problem, we present some additional results.

First, notice that, if we fix $\alpha(G)<k$, then there is a limit for $\vartheta(G)/\alpha(G)$. Specifically, the following theorem holds:

\emph{Theorem 2.} For every $k \in \mathbb{N}$ there exists an absolute constant $M_k$ such that,
for any $n$-vertex graph $G$ with $\alpha(G)< k$, $\vartheta(G) \leq M_k n^{1-2/k}$.

The proof is based on Theorem 5.1 in Ref.~\cite{AK98}, which generalizes a result in Ref.~\cite{KK83} for $k=3$, for which $M_3=2^{\frac{2}{3}}$.

Although there are no explicit constructions for general $k$, in Ref.~\cite{Alon94} there is a family of graphs
with $\alpha=2$ and $\vartheta \sim 2^{\frac{2}{3}} n^\frac{1}{3}$. These graphs depend on a parameter $r$ which cannot be a multiple of $3$.
In this case, $n=2^{3k}$. For $r=2$, it is a graph with $64$ vertices such that its complement is the graph consisting of $16$ unconnected squares. In this particular case, $\vartheta=\alpha$, therefore it is not a quantum contextuality witness. The interesting thing is that it gives us some intuition about how the graphs corresponding to AMC may be: Dense graphs (i.e., with a number of edges close to the maximal number of edges) with a very high number of vertices. We have also computed the adjacency matrix for the complement of the graph for $r=4$. It has over two million edges. We do not know whether or not the cases $r=4,5$ are quantum contextuality witnesses. However, this is the case for $r > 6$. These graphs are Cayley graphs \cite{Meier08} and, therefore, regular and vertex-transitive.

Alternatively, if we do not fix $\alpha(G)$, then we can obtain larger violations with simpler graphs
for which the number of vertices does not grow so fast.

\emph{Theorem 3.} For every $\epsilon > 0$ there is an explicit family of graphs for which $\vartheta \geq \left(\frac{1}{2} - \epsilon\right) n$ and $\alpha <n^{\delta (\epsilon)}$, $\delta (\epsilon)< 1$.

The proof is based on Theorem 6.1 in Ref.~\cite{Alon94}.

Interestingly, there are explicit quantum contextuality witnesses reaching these values. For a pair of integers $q>s>0$, $G(q,s)$ is the graph on $n=\left(\begin{array}{c} 2q\\ q\end{array}\right)$ vertices, with each vertex corresponding to a $q$-subset of $\{1,2,\ldots , 2q\}$, and such that two vertices are adjacent if and only if their intersection has exactly $s$ elements. For small values of $q$ and $s$ we have:
\begin{center}
\begin{tabular}{ccccc}
$q$ & $s$ & $n$ & $\alpha$ & $\vartheta$\\
\hline
$2$&$1$&$6$&$2$&$2$\\
$3$&$1$&$20$&$4$&$5$\\
$3$&$2$&$20$&$4$&$5$\\
$4$&$1$&$70$&$17$&$23$\\
$4$&$2$&$70$&$10$&$10$\\
$4$&$3$&$70$&$14$&$14$\\
$5$&$1$&$252$&$\geq 55$&$94.5$\\
$5$&$2$&$252$&$\geq 27$&$42$\\
$5$&$3$&$252$&$\geq 12$&$18.67$\\
$5$&$4$&$252$&$\geq 28$&$42$
\end{tabular}
\label{TableI}
\end{center}

For this family, there are explicit orthonormal representations \cite{Alon94} that achieve the lower bound on $\vartheta$ in dimension $2q$. Each of these orthonormal representation provides the measurements (the rank-one projectors onto the unit vectors of the representation) and the quantum state (the handle of the representation) for an experimental implementation of a quantum contextuality witness.


\section{Conclusions and open problems}


Contextuality is an important resource for computation and communication. So far, a given experimental scenario was said to exhibit maximal contextuality when the maximum possible violation $\vartheta$ of the corresponding noncontextual bound $\alpha$ was the maximum allowed by some principles. This approach motivated the definition of, e.g., ``fully contextual correlations'' as those in which $\vartheta$ equals the nonsignaling bound \cite{ADLPBC12}. Recent developments \cite{PPKSWZ09,NW10,Cabello13,FSABCLA12,Yan13,ATC14,Amaral14,Cabello14,Cabello15} suggest that QT could be maximally contextual in the sense that the contextuality of specific scenarios can be explained by embedding them into larger maximally contextual scenarios. However, this still does not answer the question of how large quantum contextuality can be.

Here we investigated whether QT achieves the maximum conceivable contextuality. Previous evidence, including the quantum violation of NC and Bell inequalities, strongly suggested against that possibility. Surprisingly, in QT, there are scenarios in which the ratio $\vartheta/\alpha$ can be arbitrarily close to its absolute maximum.

How this may happen? Unfortunately, our proof is not constructive and does not allow us to identify explicit scenarios. However, the examples of almost AMC that we found, when examined from the point of view of their quantum realization, have in common the presence of a very large set of rank-one projectors such that almost everyone commutes with everyone (suggesting, when coarse grained, a classical system of high dimensionality), but at the same time riddled with a large number small ``islands'' of projectors such that not all of them commute (suggesting, when fine grained, a strong quantum behavior). This supports the view that quantum AMC is an emerging phenomenon that only occurs when small quantum structures infest in a particular way seemingly classical and highly complex systems.

A natural open problem is therefore to learn more about these scenarios. On the technical side, the difficulty to identify them is related to the difficulty of identifying graphs with large $\vartheta(G)/\alpha(G)$ due to the fact that $\alpha(G)$ is hard to compute and is only known for very restricted families of graphs. A possible strategy to address this problem would be to single out graph invariants computable in polynomial time [like $\vartheta(G)$] and sandwiched between $\alpha(G)$ and $\vartheta(G)$, and then identify graphs with a large ratio between this quantity and $\vartheta(G)$. For example, in Ref.~\cite{LS79} the authors present a semidefinite program (SDP) approximation for $\alpha(G)$ which is at least as good as $\vartheta(G)$. Many approximations are also known \cite{PVZ02}, including many SDP hierarchies that converge in a finite number of steps \cite{SL96, KP02, BO04, GLRS09}. Each of them provides other SDP approximations to $\alpha(G)$, and, if smaller than $\vartheta(G)$, can be used to lower bound the contextuality. From the conceptual point of view, even if the simplest scenarios with quantum AMC are so complex that they do not allow for experimental tests using present technologies, it seems to be important to understand how AMC emerges, what implications AMC has (in particular, for the limits of the contextuality of simple scenarios), and what AMC may be useful for.

Finally, another important question is which is the maximum nonlocality allowed by QT. This problem has been only studied for fixed scenarios and has been proven hard even in the simplest scenarios. The quest for maximum nonlocality with no restriction on the scenario is much harder. However, the same way the graph-theoretic approach to quantum contextuality \cite{CSW10,CSW14,AFLS12} has been useful to answer the question of what is the maximum contextuality in QT, the multigraph approach of Ref.~\cite{RDLTC14} may help to address the question of which is the maximum nonlocality in QT.


\begin{acknowledgments}
The authors thank Israel Vainsencher for his help with the graph of Theorem~2 and Gustavo Ca\~{n}as for his help with Fig.~\ref{Fig1}. This work was supported by the FQXi large grant project ``The Nature of Information in Sequential Quantum Measurements,'' Projects No.\ FIS2011-29400 and No.\ FIS2014-60843 (MINECO, Spain) with FEDER funds, the program ``Science without Borders'' (CAPES and CNPq, Brazil), CAPES, CNPq, and FAPEMIG (Brazil). Part of this work was done in the Centro de Ciencias de Benasque ``Pedro Pascual.''
\end{acknowledgments}

%
%

\end{document}